\documentclass[aps,prb,letter,preprint,superscriptaddress]{revtex4-1}

\usepackage{hyperref}

\usepackage{srcltx}
\usepackage{psfrag}
\usepackage{epsfig}
\usepackage{color}
\usepackage[tbtags,sumlimits,nointlimits,reqno]{amsmath}
\usepackage{amssymb}
\usepackage{float}
\usepackage{subfigure}
\usepackage{arydshln}

\begin{document}

\title{Terahertz Magnetoplasmon Energy Concentration and Splitting\\ in Graphene PN Junctions}

\author{Nima Chamanara}
\affiliation{Poly-Grames Research Center, \'{E}cole Polytechnique de Montr\'{e}al, Montr\'{e}al, Qu\'{e}bec H3T 1J4, Canada.}
\author{Dimitrios L. Sounas}
\affiliation{Department of Electrical and Computer Engineering, The University of Texas at Austin, Austin, Texas 78712, USA.}
\author{Thomas Szkopek}
\affiliation{Department of Electrical and Computer Engineering, McGill University, Montr\'{e}al, Qu\'{e}bec H3A 2A7, Canada.}
\author{Christophe Caloz}
\affiliation{Poly-Grames Research Center, \'{E}cole Polytechnique de Montr\'{e}al, Montr\'{e}al, Qu\'{e}bec H3T 1J4, Canada.}

\begin{abstract}
Terahertz plasmons and magnetoplasmons propagating along electrically and chemically doped graphene p-n junctions are investigated. It is shown that such junctions support non-reciprocal magnetoplasmonic modes which get concentrated at the middle of the junction in one direction and split away from the middle of the junction in the other direction under the application of an external static magnetic field. This phenomenon follows from the combined effects of circular birefringence and carrier density non-uniformity. It can be exploited for the realization of plasmonic isolators.
\end{abstract}

\maketitle

\section{Introduction}

The linear band structure, tunability and ambipolarity of graphene \cite{Novoselov2004,Geim2007} have recently opened up new horizons in the area of plasmonics. These fundamental properties lead to unique plasmonic phenomena, such as the existence of both TE and TM plasmons~\cite{Grigorenko2012,Mikhailov2007,Hanson2008,Crassee2012,Vakil2011Sci} and voltage tunable plasmonic modes \cite{Silvestrov2008,Mishchenko2012,thongrattanasiri2012,Petkovi2013}. These modes have been recently investigated towards the realization of various enhanced plasmonic devices \cite{Grigorenko2012,Echtermeyer2011,Mueller2010,Gabor04112011,Vakil2011Sci,thongrattanasiri2012,Chamanara_coupler_2013}.

Using electrical gating, one can modify and tune the charge profile to generate useful modes on graphene structures. For instance, one can obtain regions with opposite carrier types on a graphene strip by applying a tangential transverse electric field across it so as to create a p-n junction. In~\cite{Mishchenko2012}, it was theoretically shown at zero temperature that such a junction supports localized plasmonic and magnetoplasmonic modes with the magnetoplasmonic modes existing only for one direction of propagation (the modes in the other direction being cut off) in the limit of a very high magnetic bias field.

Allowing non-zero temperatures leads to the excitation of thermal carriers, which induce non-zero conductivity at the center of the junction and hence affects the electromagnetic field distribution across the strip. Moreover, allowing arbitrary magnetic bias fields leads to the existence of a circular birefringence regime where magnetoplasmonic modes exist in the two propagation directions but exhibit very different distributions, one being concentrated  at the center of the junction and the other one being split away from it. This paper reveals these non-reciprocal phenomena and proposes isolator devices based on them.

\begin{figure}[ht!]
\begin{center}
\subfigure[]{\label{fig:GrapheneP-N-Junctions_electrical}
\psfrag{B}[c][c][0.8]{$B_0$}
\psfrag{E}[c][c][0.8]{$E_0$}
\psfrag{z}[c][c][0.8]{$z$}
\psfrag{x}[c][c][0.8]{$x$}
\psfrag{w}[c][c][0.8]{$w$}
\includegraphics[width=0.4\columnwidth]{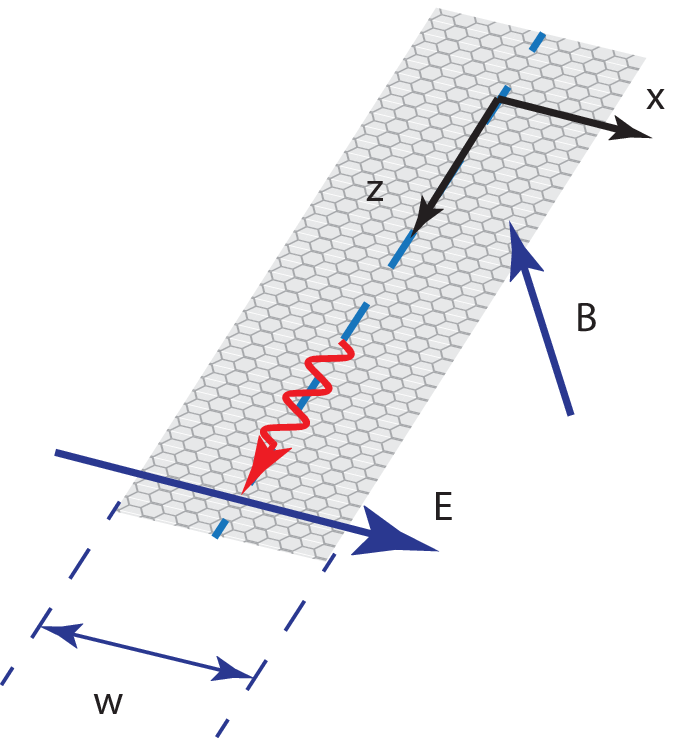}}
\subfigure[]{\label{fig:GrapheneP-N-Junctions_chemical}
\psfrag{B}[c][c][0.8]{$B_0$}
\psfrag{z}[c][c][0.8]{$z$}
\psfrag{x}[c][c][0.8]{$x$}
\psfrag{w}[c][c][0.8]{$w$}
\psfrag{s}[c][c][0.8]{$s$}
\psfrag{n}[c][c][0.9]{n doped}
\psfrag{p}[c][c][0.9]{p doped}
\includegraphics[width=0.45\columnwidth]{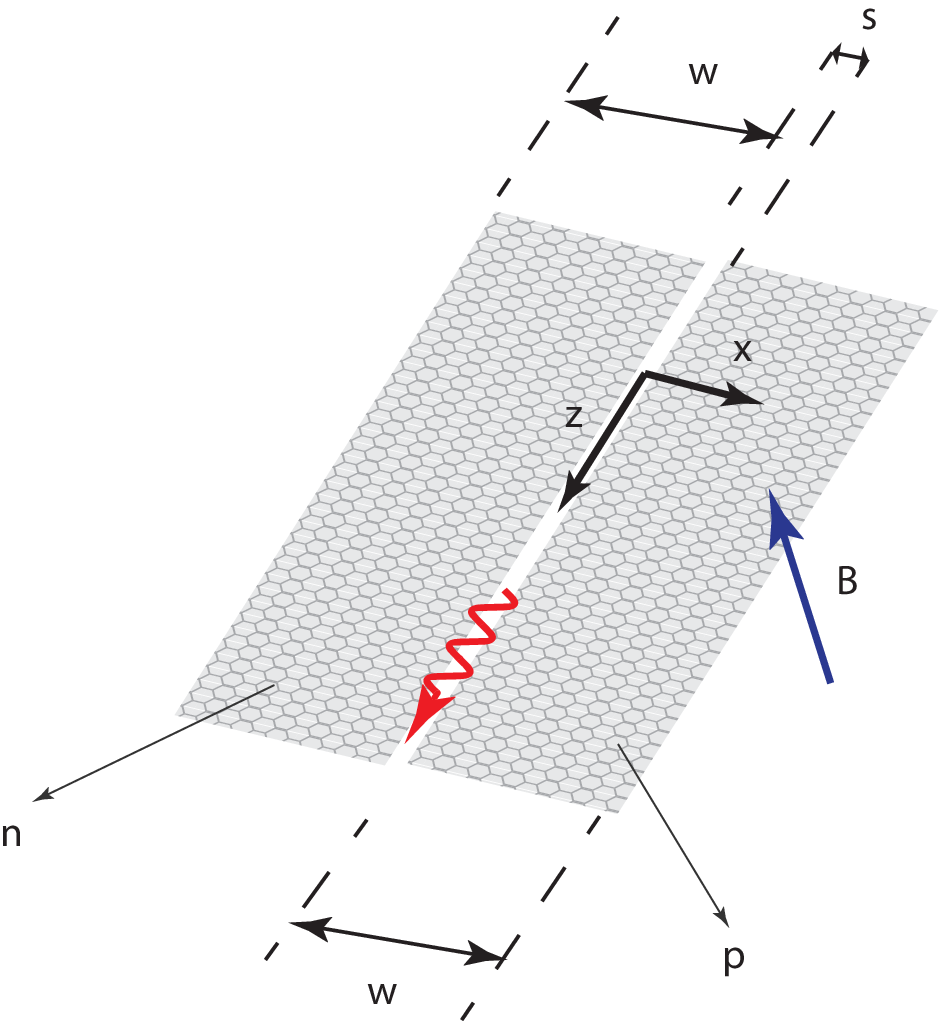}}
\caption{(a) Electrically and (b) chemically doped graphene p-n junctions.}
\label{fig:GrapheneP-N-Junctions}
\end{center}
\end{figure}

The magnetoplasmonic modes are computed by numerically solving Maxwell equations for the graphene strip with exact non-uniform carrier densities in the Kubo conductivity, which properly captures the quantized Landau regime prevailing near the center of the junction. Two types of doping are considered and compared: doping by a transverse electric field and chemical doping, represented in Figs.~\ref{fig:GrapheneP-N-Junctions_electrical} and~\ref{fig:GrapheneP-N-Junctions_chemical}, respectively. The corresponding structures are analyzed with the 2D finite difference frequency domain technique (FDFD)~\cite{YZhao2002}, where graphene is modeled as a zero-thickness conductive sheet with a conductivity tensor given by the Kubo formula~\cite{Gusynin2007,Gusynin2009}.

\section{Magnetoplasmon Energy Concentration and Splitting in Graphene p-n Junctions} \label{sec:ElectricPNJunction}
\subsection{Electrically Doped Graphene}

\begin{figure}[ht!]
\begin{center}
\psfrag{x}[c][c][0.8]{$x$}
\psfrag{z}[c][c][0.8]{$z$}
\psfrag{w}[c][c][0.8]{$w$}
\psfrag{e}[c][c][0.8]{$E_0$}
\psfrag{l}[c][c]{$2x/w$}
\psfrag{y}[c][l][0.9][90]{net carrier density (cm$^{-2}$)}
\psfrag{r}[c][c][0.9][90]{potential (V)}
\psfrag{n}[r][c][0.9]{$n_{\text{net}}=n-p$}
\includegraphics[width=1.0\columnwidth]{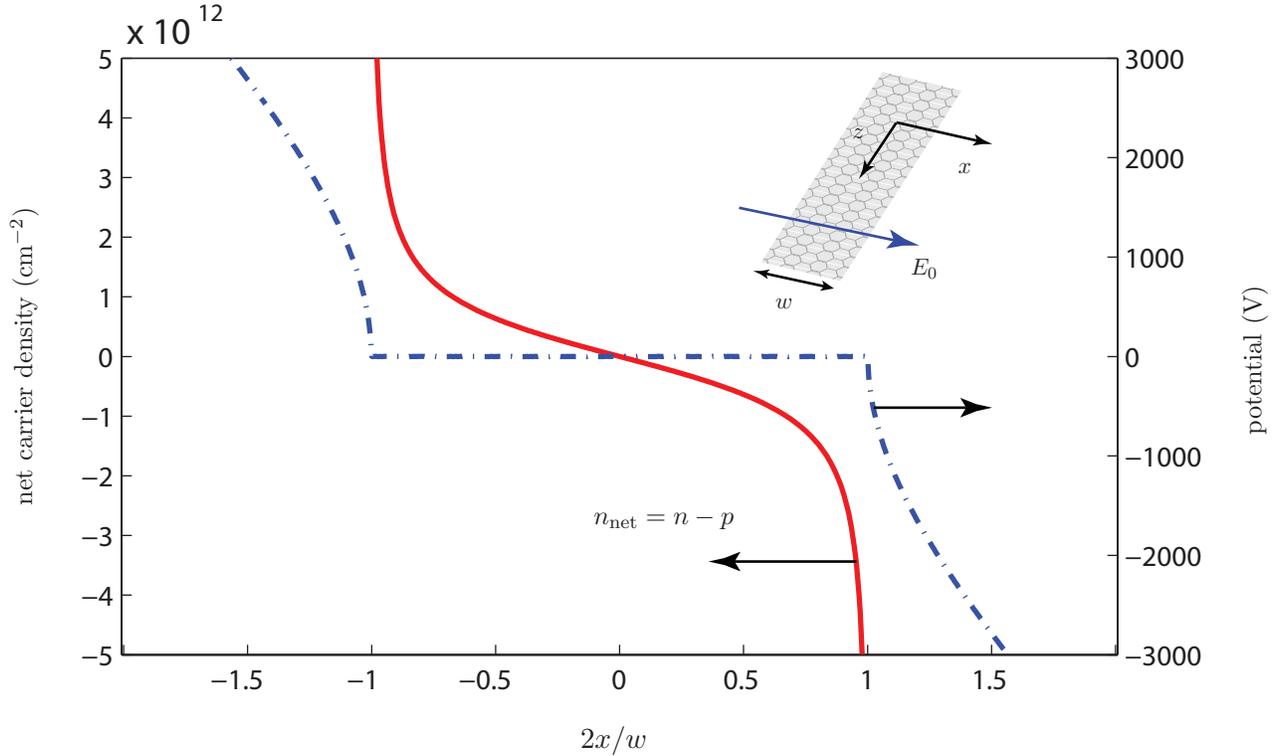}
\caption{Net carrier density and electric potential for a graphene strip doped with an electric field; \mbox{$w=50$~$\mu$m} and \mbox{$E_0=10^8$~V/m}.}
\label{fig:ElectricFieldGatedGraphene}
\end{center}
\end{figure}

First consider the electrically doped graphene structure [Fig.\ref{fig:GrapheneP-N-Junctions_electrical}]. A transverse electric field applied tangentially to the graphene strip creates a non-uniform carrier density and therefore a non-uniform conductivity across the strip. The net charge profile can be found by solving the integral equation

\begin{subequations}
\begin{equation} \label{eq:carrier_density_doping_with_electric_field}
\int_{-w/2}^{w/2} \rho(x',y')G(x,y;x',y')dx' -E_0 x = 0,
\end{equation}
with
\begin{equation}
-w/2 \leq x\leq w/2, \text{~~}y=0, \text{~~}y'=0,
\end{equation}
\end{subequations}

\noindent for the net charge density $\rho(x',y')$, where $G(x,y;x',y')=\dfrac{-1}{2\pi\epsilon_0}\ln\sqrt{(x'-x)^2+(y'-y)^2}$ is the 2D free-space Green function for the Poisson equation and $E_0$ is the applied electrostatic field. The resulting net carrier density, $n_{\text{net}}=n-p$, where $n$ and $p$ are the electron and hole densities, respectively, is plotted in Fig.~\ref{fig:ElectricFieldGatedGraphene} along with the electric potential for a graphene strip of width $w=50$~$\mu$m. Different carrier types symmetrically appear at the opposite sides of the strip, which results in the formation of a p-n junction at the center of the strip.

\begin{figure}[ht!]
\begin{center}
\psfrag{x}[c][c][0.9]{$2x/w$}
\psfrag{y}[c][l][0.9][90]{carrier density (cm$^{-2}$)}
\psfrag{a}[c][c][1.0]{$p$}
\psfrag{b}[c][c][1.0]{$n$}
\psfrag{c}[l][c][1.0]{$|n_{\text{net}}|$}
\psfrag{d}[c][c][1.0]{$n+p$}
\includegraphics[width=1.0\columnwidth]{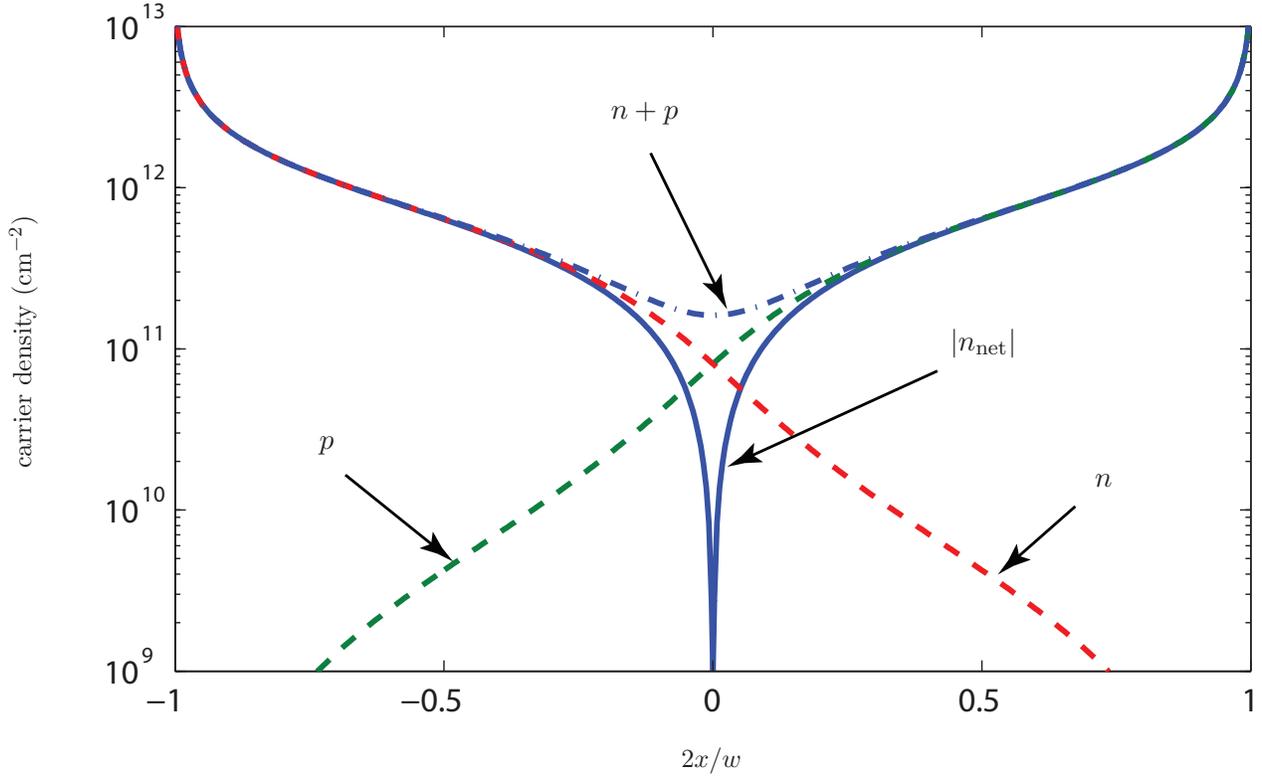}
\caption{Thermally excited electron and hole densities at room temperature for a graphene strip doped with an electric field; \mbox{$w=50$~$\mu$m}, \mbox{$E_0=10^8$~V/m}, \mbox{$T=300$~K}.}
\label{fig:ElectronHole_Density}
\end{center}
\end{figure}

\begin{figure}[ht!]
\begin{center}
\psfrag{x}[c][c][0.9]{$2x/w$}
\psfrag{y}[c][l][0.9][90]{chemical potential~(eV)}
\psfrag{z}[c][c][0.9][90]{conductivity~(mS)}
\psfrag{r}[l][c][1.0]{real}
\psfrag{l}[l][c][1.0]{imag}
\includegraphics[width=1.0\columnwidth]{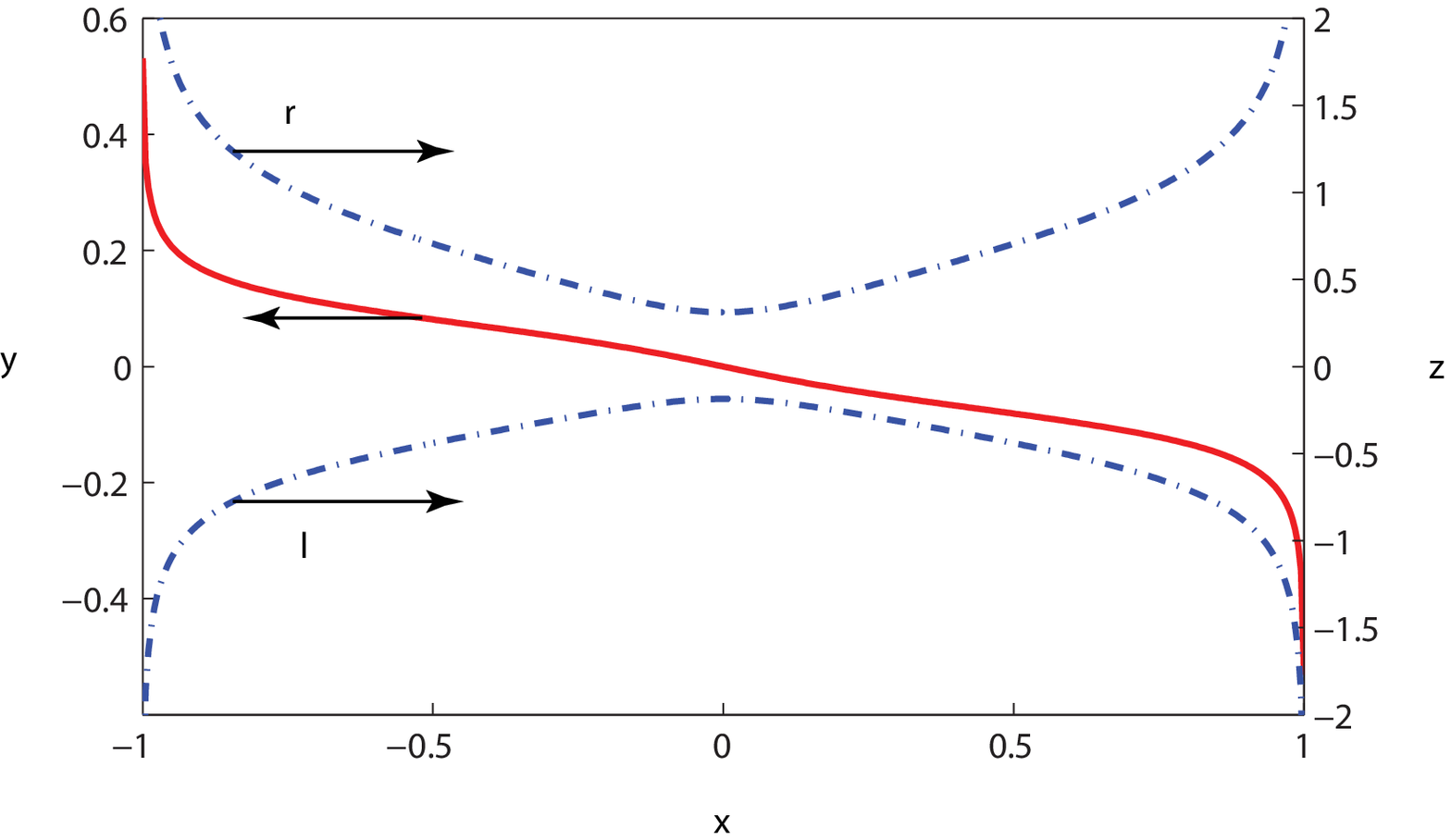}
\caption{Chemical potential and Kubo conductivity for a graphene strip doped with an electric field; \mbox{$w=50$~$\mu$m} and \mbox{$E_0=10^8$~V/m}, \mbox{$B_0=0$~T}, \mbox{$\tau=0.1$~ps}, \mbox{$T=300$~K}, \mbox{$f=1$~THz}.}
\label{fig:chemical_pot_conductivity}
\end{center}
\end{figure}

The chemical potential ($\mu_c$) and the electron and hole densities are found from $n_{\text{net}}$ by numerically solving the equation

\begin{equation} \label{eq:net_density}
n_{\text{net}} = \int_{0}^{\infty} f_d(E,\mu_c)N(E)dE -\int_{-\infty}^{0} [1-f_d(E,\mu_c)]N(E)dE,
\end{equation}

\noindent
where $f_d(E,\mu_c)$ is the Fermi-Dirac distribution and $N(E)=\frac{2|E|}{\pi\hbar^2 v_f^2}$ is the density of energy states. The integrals represent the electron and hole densities ($n$ and $p$, respectively). The carrier densities are plotted in Fig.~\ref{fig:ElectronHole_Density} while the corresponding chemical potential and Kubo conductivity are plotted in Fig.~\ref{fig:chemical_pot_conductivity}. Note that at the center of the strip, despite the zero net carrier density, a significant amount of thermally excited electrons and holes are present, which leads to the significant conductivity observed in Fig.~\ref{fig:chemical_pot_conductivity}. For simplicity, we assumed an energy independent scattering time of $\tau=0.1$~ps. Moreover, we assumed that the graphene has less than $10^{10}$~cm$^{-2}$ of unintentional doping fluctuations.

\begin{figure}[ht!]
\begin{center}
\subfigure{\label{fig:ElectricFieldGatedGraphenePlasmon_a}
\psfrag{x}[c][c][0.9]{$\text{Re}(k_z/k_0)$}
\psfrag{y}[c][c][0.9][90]{$f$~(THz)}
\psfrag{i}[l][c][0.9]{mode~1}
\psfrag{j}[l][c][0.9]{mode~2}
\psfrag{k}[c][c][0.9]{mode~3}
\psfrag{l}[l][c][0.9]{mode~4}
\includegraphics[width=1.0\columnwidth]{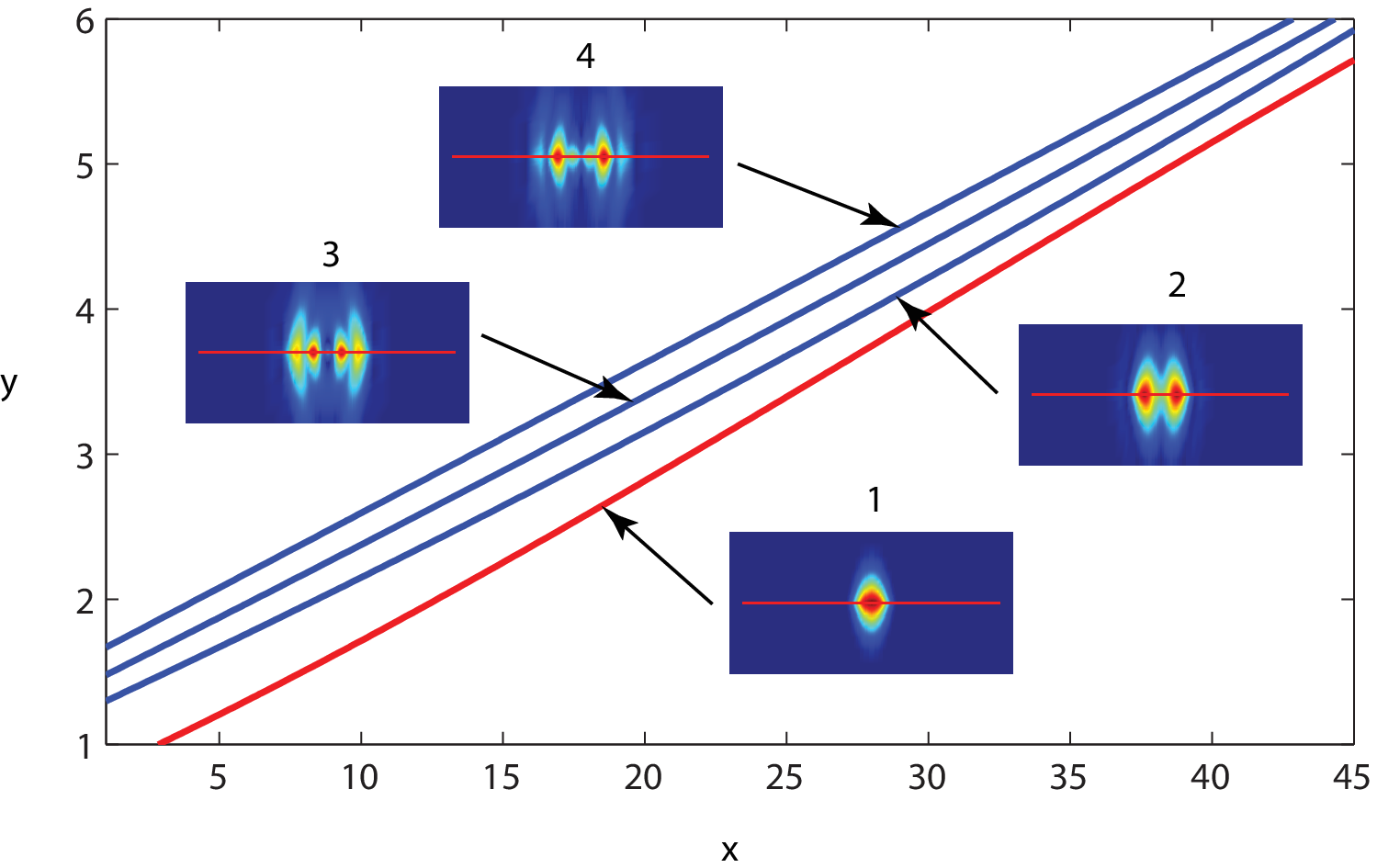}}
\subfigure{\label{fig:ElectricFieldGatedGraphenePlasmon_b}
\psfrag{x}[c][c][0.9]{$-\text{Im}(k_z/k_0)$}
\psfrag{y}[c][c][0.9][90]{$f$~(THz)}
\includegraphics[width=1.0\columnwidth]{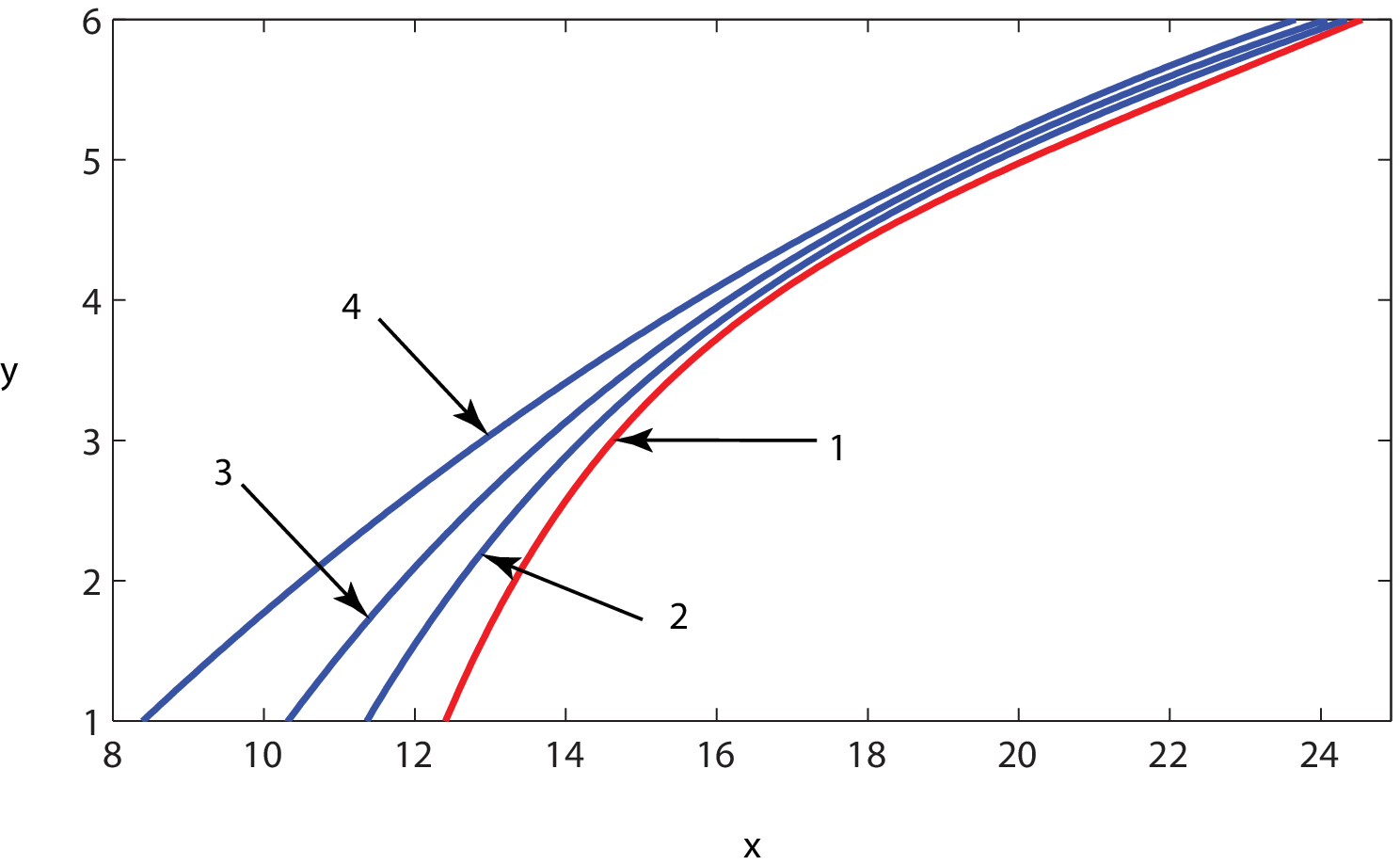}}
\caption{Slow-wave factor and loss for a graphene strip biased by an electric field; \mbox{$w=50$~$\mu$m}, \mbox{$E_0=10^8$~V/m}, \mbox{$B_0=0$~T}, \mbox{$\tau=0.1$~ps}, \mbox{$T=300$~K}. The p-n junction mode is represented in red.}
\label{fig:ElectricFieldGatedGraphenePlasmon}
\end{center}
\end{figure}

The graphene strip is next simulated as a conductive sheet in the FDFD resolution of Maxwell equations in order to compute the plasmonic and magnetoplasmonic modes. The slow-wave factor and loss for the plasmon modes propagating along the structure are shown in Fig.~\ref{fig:ElectricFieldGatedGraphenePlasmon}, where $k_z$ and $k_0$ are the propagation constant in the $z$ direction and the free space wave number, respectively. The structure supports a plasmon mode localized at the p-n junction, an infinite number of bulk modes (only the first four are shown here) and two edge modes (not shown here). The losses for the first few modes are shown in Fig.~\ref{fig:ElectricFieldGatedGraphenePlasmon_b}. Since the carrier density is low near the center of the graphene strip, the p-n junction mode has higher loss than the other modes. The farther the mode is from the center, the higher its loss is.

\begin{figure}[ht!]
\begin{center}
\subfigure{\label{fig:ElectricFieldGatedGrapheneMagnetoPlasmon_a}
\psfrag{x}[c][c][0.9]{$\text{Re}(k_z/k_0)$}
\psfrag{y}[c][c][0.9][90]{$f$~(THz)}
\psfrag{r}[c][c]{mode~$1^{+}$}
\psfrag{l}[c][c]{mode~$1^{-}$}
\includegraphics[width=1.0\columnwidth]{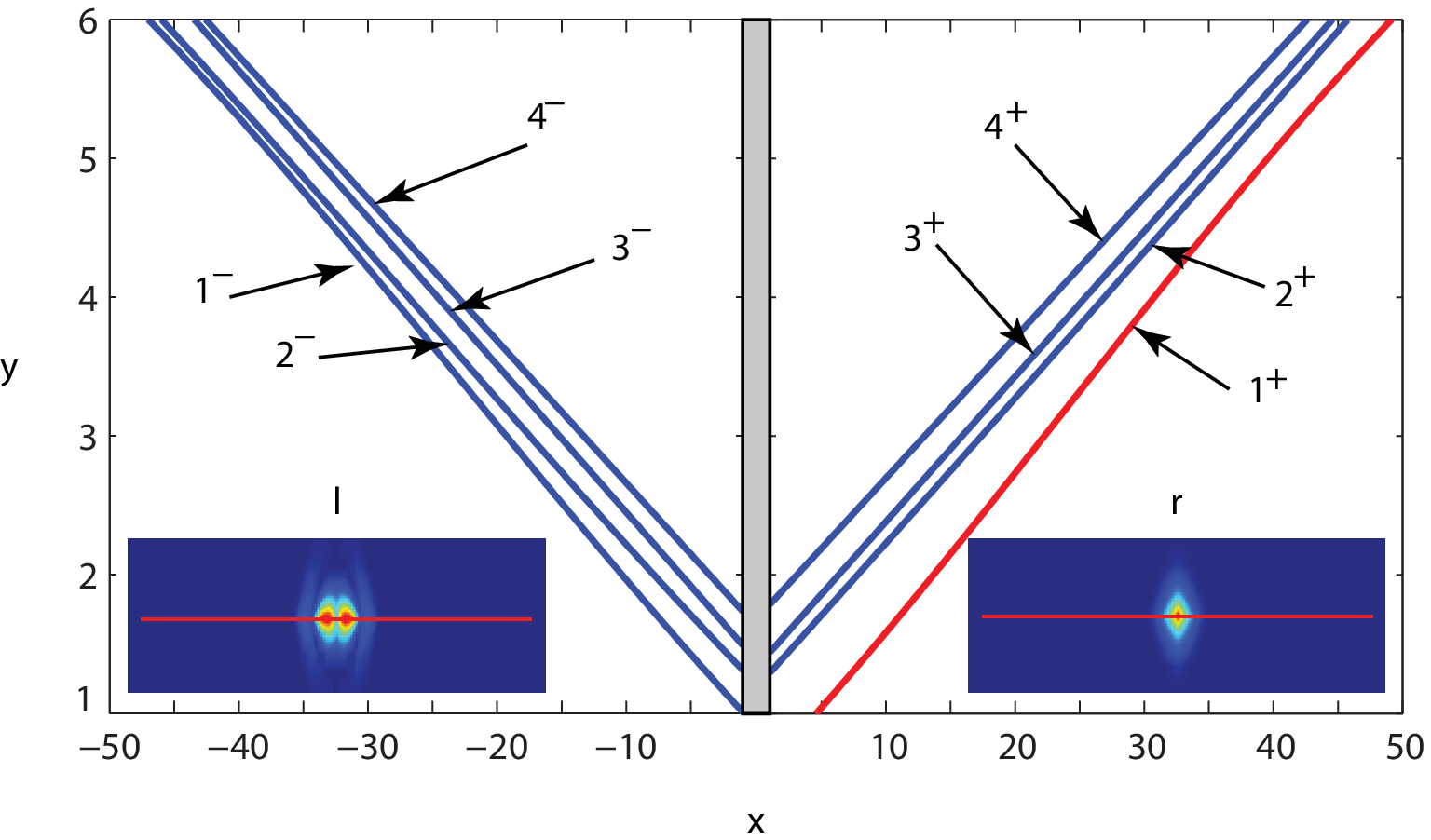}}
\subfigure{\label{fig:ElectricFieldGatedGrapheneMagnetoPlasmon_b}
\psfrag{x}[c][c][0.9]{$-\text{Im}(k_z/k_0)$}
\psfrag{y}[c][c][0.9][90]{$f$~(THz)}
\includegraphics[width=1.0\columnwidth]{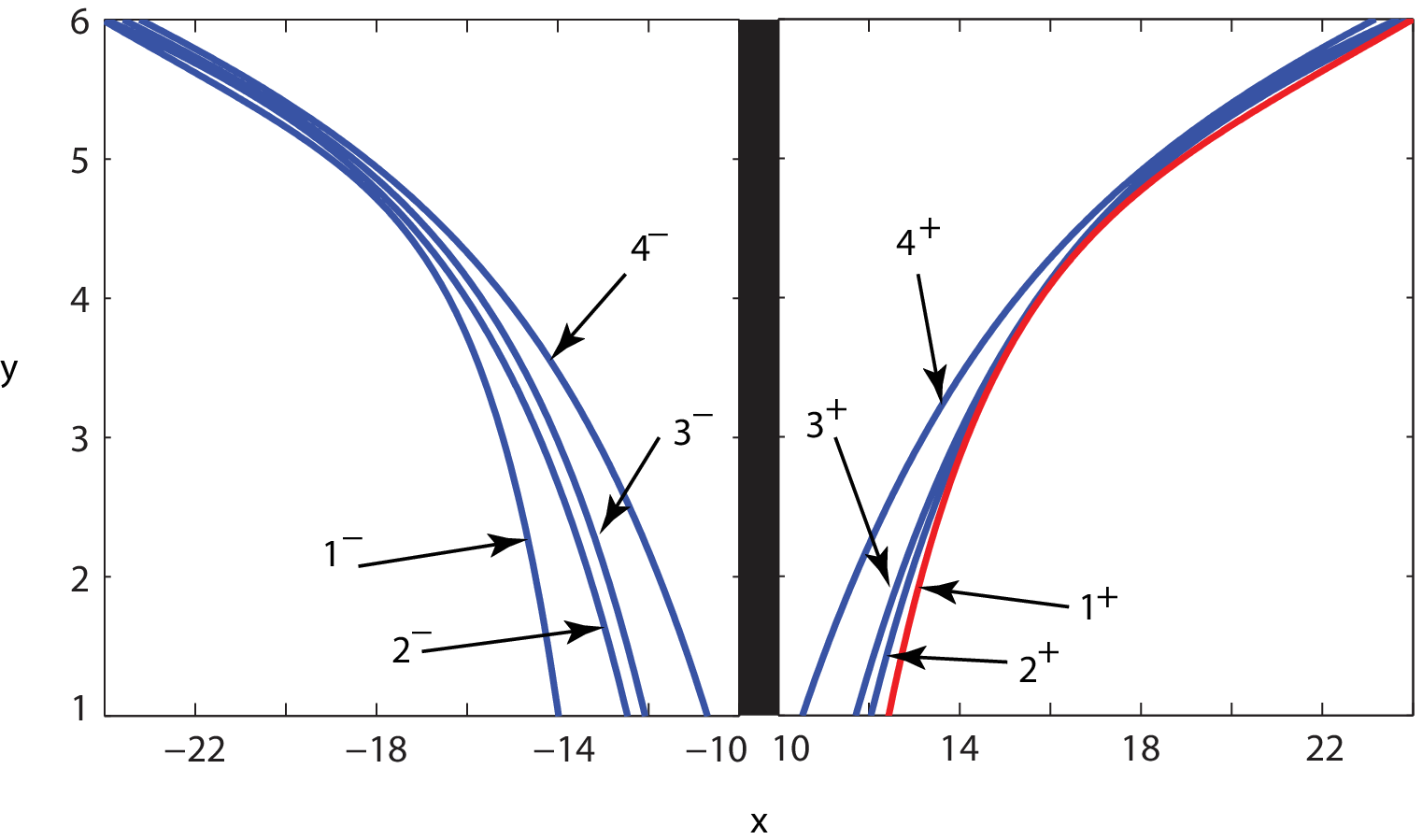}}
\caption{Dispersion curves for a magnetically biased graphene strip biased by an electric field; \mbox{$w=50$~$\mu$m}, \mbox{$E_0=10^8$~V/m}, \mbox{$B_0=0.1$~T}, \mbox{$\tau=0.1$~ps}, \mbox{$T=300$~K}. The p-n junction mode is represented in red. The grey area represents the light cone.}
\label{fig:ElectricFieldGatedGrapheneMagnetoPlasmon}
\end{center}
\end{figure}

In the presence of a magnetic biasing field, the lowest magnetoplasmon mode exhibits particularly interesting non-reciprocal properties. If the field is sufficiently high, it propagates only in one direction, as shown in the dispersion diagram of Fig.~\ref{fig:ElectricFieldGatedGrapheneMagnetoPlasmon} for $B_0=0.1$~T. As the magnetic field is increased, the forward mode concentrates at the center, whereas its energy splits away from the center in the backward direction, as shown in Fig.~\ref{fig:ElectricFieldGatedGrapheneMagnetoPlasmon}. Therefore, if the strip is excited at its center, in the forward direction, the mode whose energy is localized at the junction is excited, whereas no modes are excited in the backward direction, as there is no mode with energy at the center. However, the source beam will need to be highly confined to avoid exciting the backward mode. Note that the modes exhibit non-commensurate field patterns, with energy being squeezed near the center, as a result of the non-uniform conductivity. In the chemically doped p-n junction, to be studied next, it will be shown that the higher order modes exhibit commensurate resonances with higher contrast between the field patterns of the p-n junction mode in the forward and backward direction.

\subsection{Chemically Doped Graphene}

\begin{figure}[ht!]
\begin{center}
\subfigure{\label{fig:PN_isolator_unbiased_dispersion_a}
\psfrag{x}[c][c][0.9]{$\text{Re}(k_z/k_0)$}
\psfrag{y}[c][c][0.9][90]{$f$~(THz)}
\includegraphics[width=1.0\columnwidth]{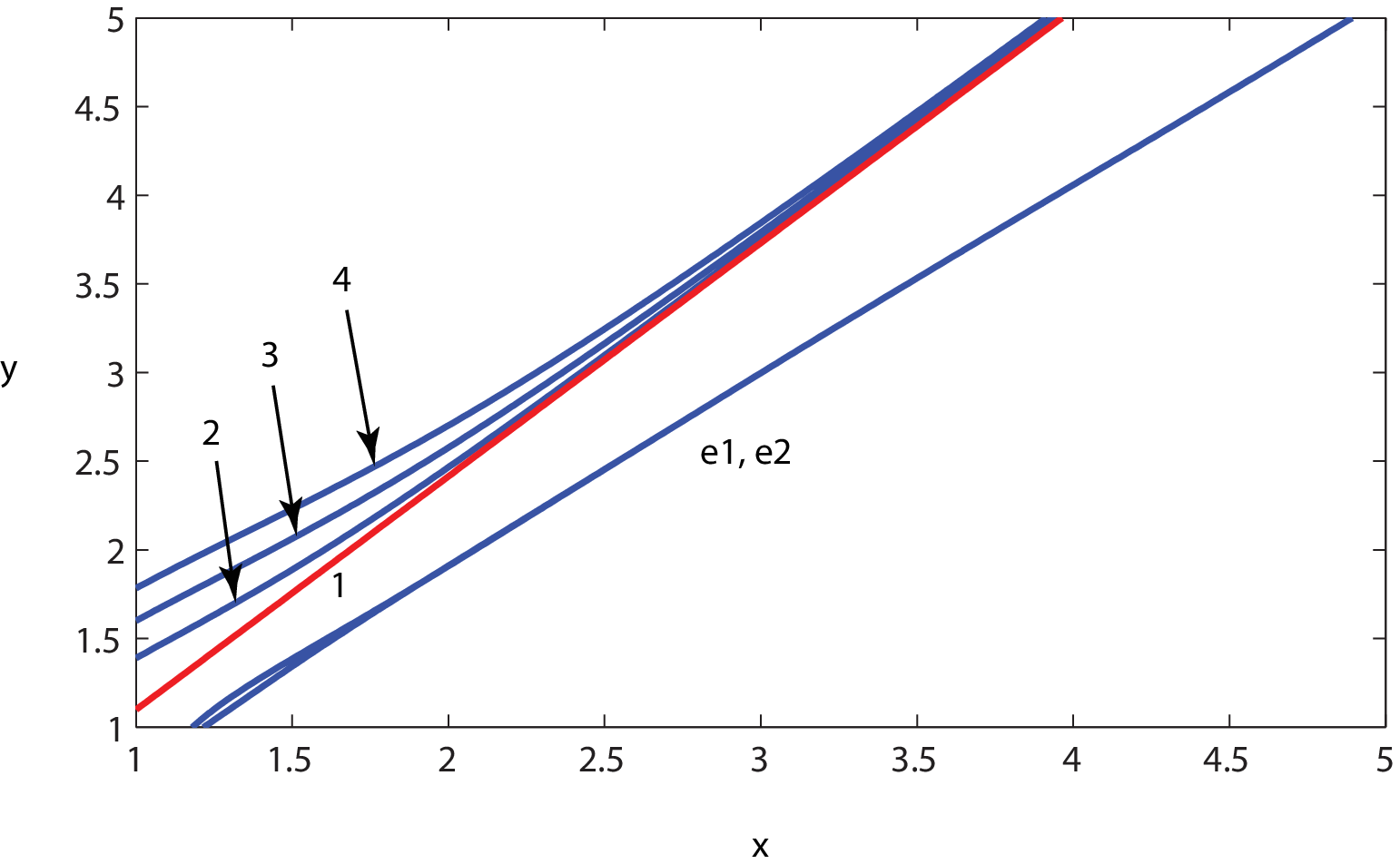}}
\subfigure{\label{fig:PN_isolator_unbiased_dispersion_b}
\psfrag{x}[c][c][0.9]{$-\text{Im}(k_z/k_0)$}
\psfrag{y}[c][c][0.9][90]{$f$~(THz)}
\includegraphics[width=1.0\columnwidth]{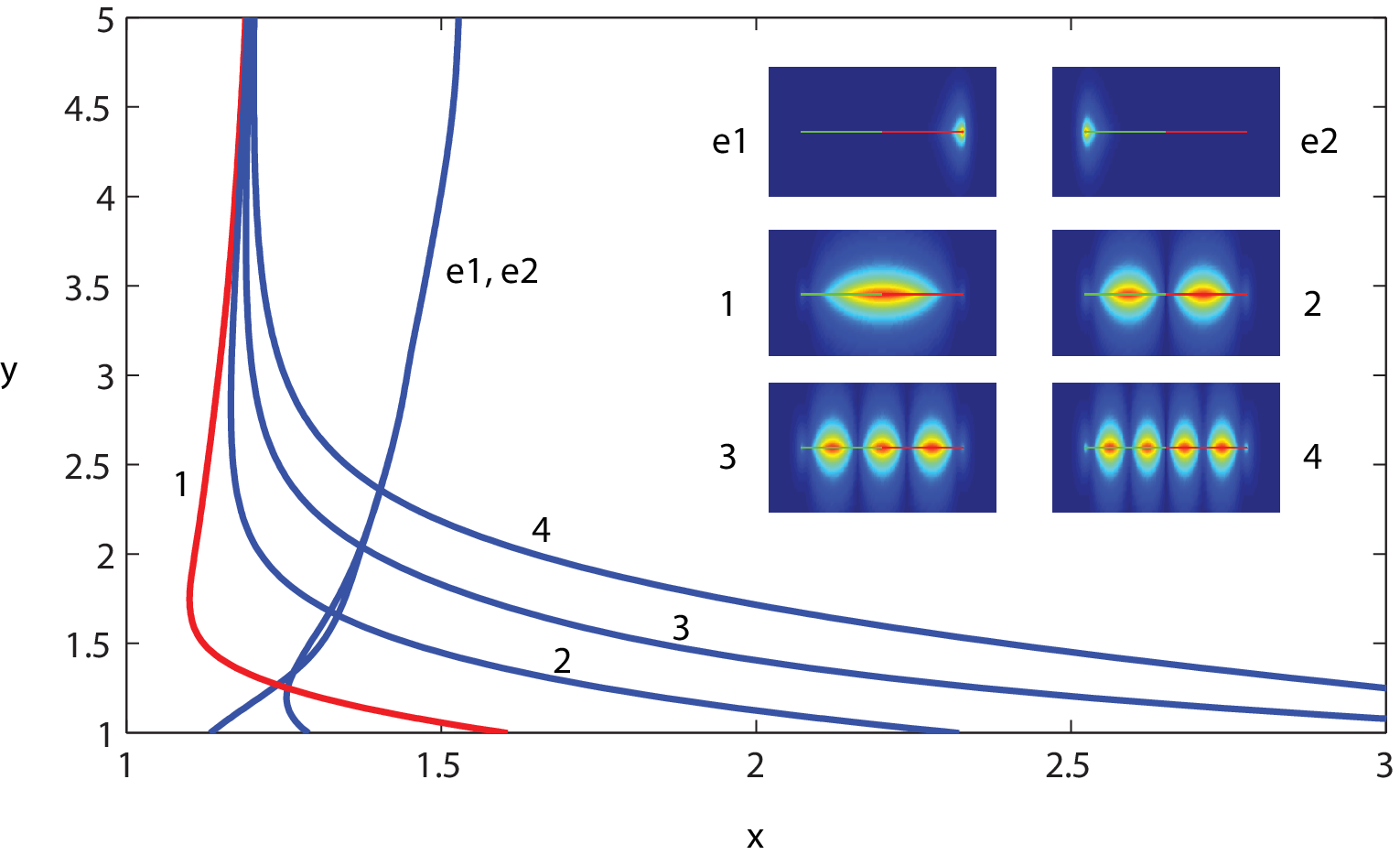}}
\caption{Slow-wave factor and loss for the structure in Fig.~\ref{fig:GrapheneP-N-Junctions_chemical} with no magnetic bias; \mbox{$w=100$~$\mu$m}, \mbox{$s=10$~nm}, \mbox{$n=p=10^{13}$~cm$^{-2}$}, \mbox{$B_0=0$~T}, \mbox{$\tau=0.1$~ps}, \mbox{$T=300$~K}. The p-n junction mode is represented in red.}
\label{fig:PN_isolator_unbiased_dispersion}
\end{center}
\end{figure}

\begin{figure}[ht!]
\begin{center}
\subfigure{\label{fig:PN_isolator_biased_dispersion_a}
\psfrag{x}[c][c][0.9]{$\text{Re}(k_z/k_0)$}
\psfrag{y}[c][c][0.9][90]{$f$~(THz)}
\psfrag{r}[c][c]{mode~$1^{+}$}
\psfrag{l}[c][c]{mode~$1^{-}$}
\includegraphics[width=1.0\columnwidth]{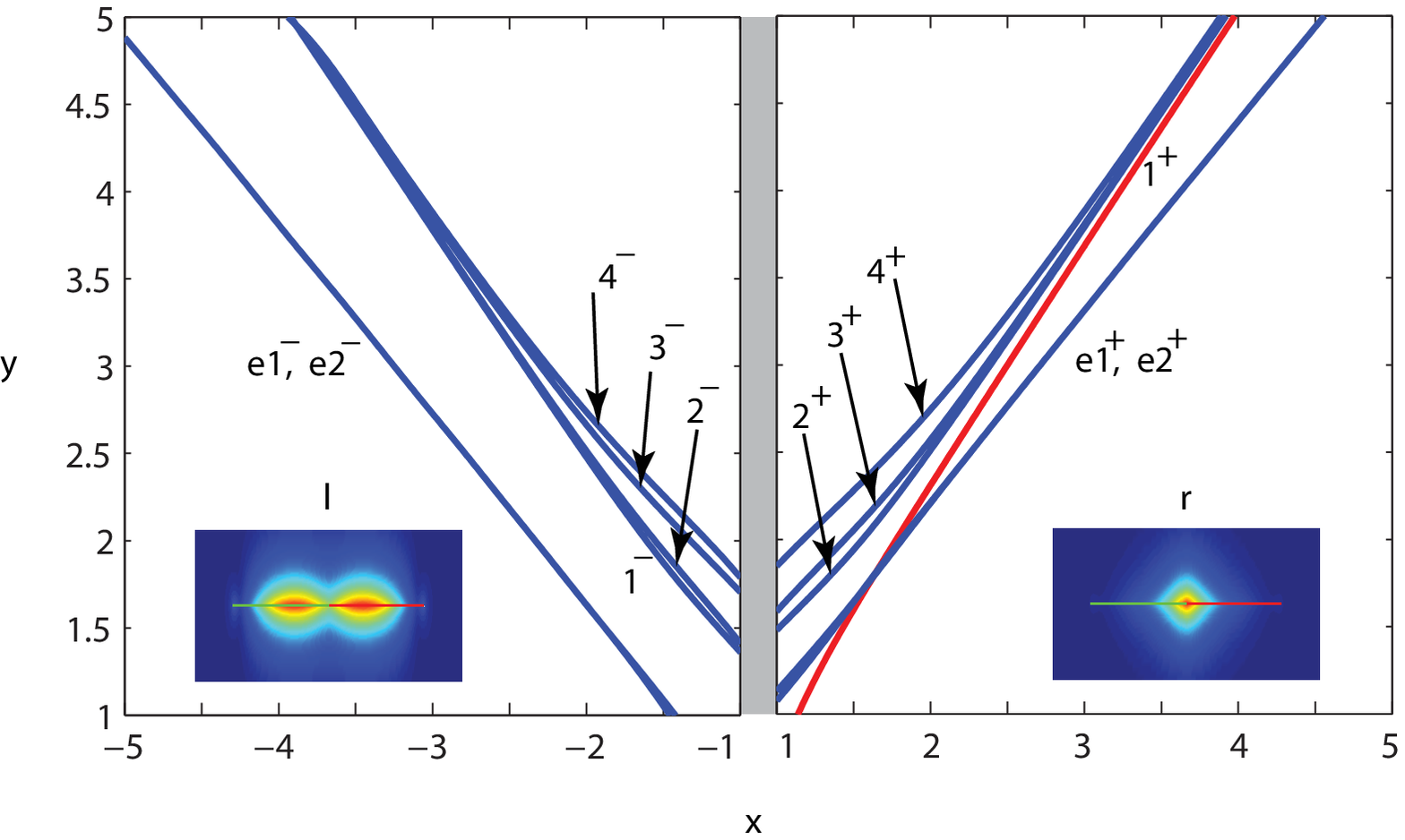}}
\subfigure{\label{fig:PN_isolator_biased_dispersion_b}
\psfrag{x}[c][c][0.9]{$-\text{Im}(k_z/k_0)$}
\psfrag{y}[c][c][0.9][90]{$f$~(THz)}
\includegraphics[width=1.0\columnwidth]{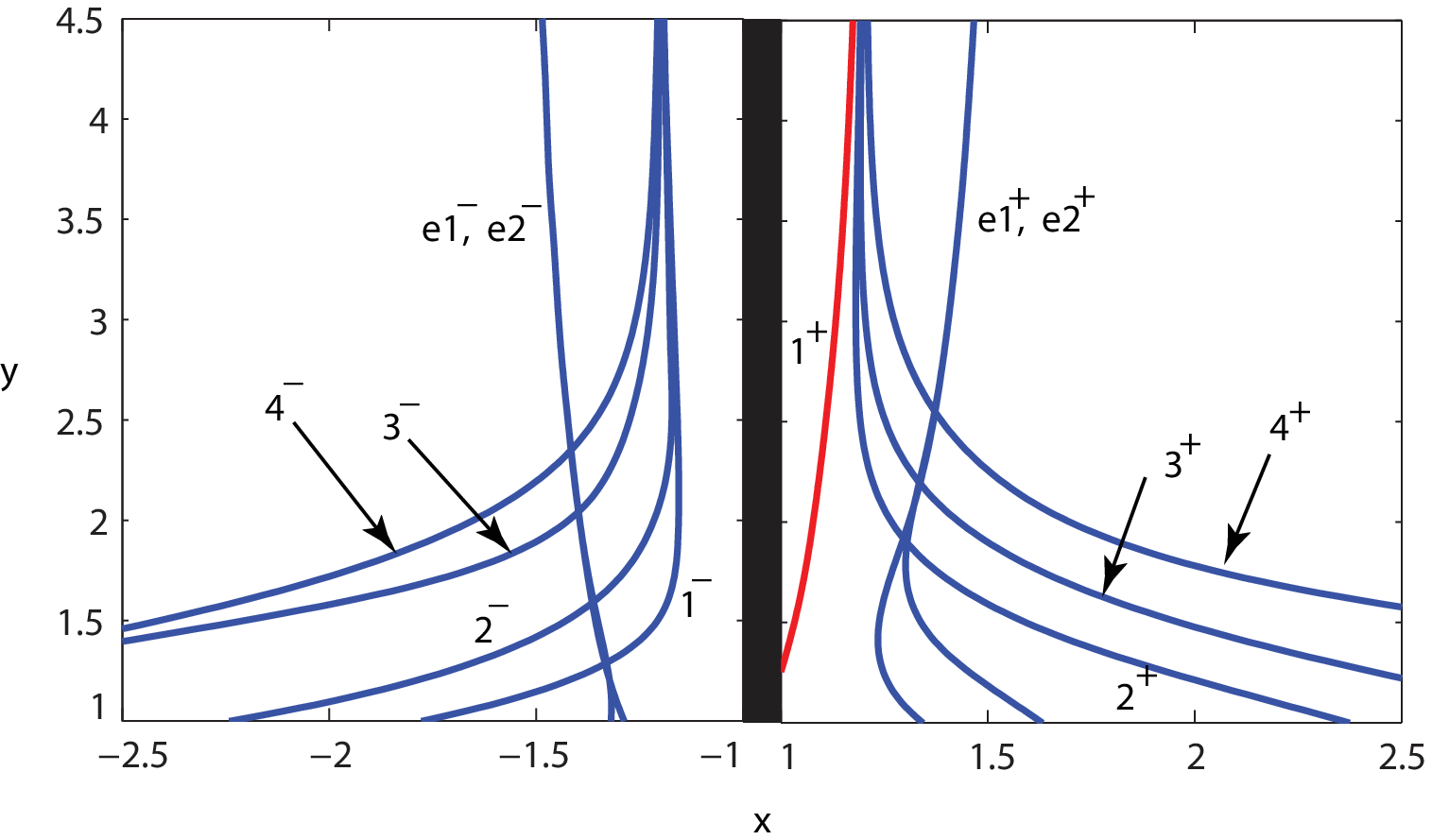}}
\caption{Slow-wave factor and loss for the magnetoplasmonic isolator in Fig.~\ref{fig:GrapheneP-N-Junctions_chemical}; \mbox{$w=100$~$\mu$m}, \mbox{$s=10$~nm}, \mbox{$n=p=10^{13}$~cm$^{-2}$}, \mbox{$B_0=1$~T}, \mbox{$\tau=0.1$~ps}, \mbox{$T=300$~K}. The p-n junction mode is represented in red. The grey area represents the light cone.}
\label{fig:PN_isolator_biased_dispersion}
\end{center}
\end{figure}

The chemically doped p-n junction is shown in Fig.~\ref{fig:GrapheneP-N-Junctions_chemical}. This structure is composed of two chemically doped graphene strips with opposite polarities forming a p-n junction and separated by a nano-gap isolating electron and hole carriers. As the structure of Fig.~\ref{fig:GrapheneP-N-Junctions_electrical}, this structure supports a plasmonic mode localized at the middle of the p-n junction. The dispersion curves for the non-biased structure is shown in Fig.~\ref{fig:PN_isolator_unbiased_dispersion}. The structure supports two edge modes, an infinite number of bulk modes, and the p-n junction mode, plotted in red. The unbiased structure has symmetric dispersion for the forward and backward propagation directions. However, as the magnetic bias is applied, time reversal symmetry is broken and the p-n junction mode exhibits different properties for the forward and backward directions.
The dispersion curves for the structure of Fig.~\ref{fig:GrapheneP-N-Junctions_chemical} under magnetic bias is plotted in Fig.~\ref{fig:PN_isolator_biased_dispersion} for a magnetic bias of $B_0=1$~T. It is seen that the mode propagating at the junction exhibits very different properties for the forward and backward directions. In the forward direction, this mode is concentrated at the center, whereas in the backward direction it has very little energy at the center.


%

%


\section{Phenomenological Explanation}

\begin{figure}[ht!]
\begin{center}
\psfrag{f}[c][c][0.9]{forward}
\psfrag{b}[c][c][0.9]{backward}
\psfrag{r}[c][c][0.9]{$R$}
\psfrag{l}[c][c][0.9]{$L$}
\psfrag{p}[c][c][0.9][90]{p doped}
\psfrag{n}[c][c][0.9][90]{n doped}
\psfrag{h}[r][c][0.9]{$\sigma = \sigma_d + j\sigma_{on}$}
\psfrag{i}[l][c][0.9]{$\sigma = \sigma_d - j\sigma_{op}$}
\psfrag{j}[r][c][0.9]{$\sigma = \sigma_d - j\sigma_{on}$}
\psfrag{k}[l][c][0.9]{$\sigma = \sigma_d + j\sigma_{op}$}
\psfrag{s}[r][c][0.9]{$=\sigma_d + j\sigma_{o}~~$}
\psfrag{t}[l][c][0.9]{$~~=\sigma_d + j\sigma_{o}$}
\psfrag{u}[r][c][0.9]{$=\sigma_d - j\sigma_{o}~~$}
\psfrag{v}[l][c][0.9]{$~~=\sigma_d - j\sigma_{o}$}
\includegraphics[width=0.8\columnwidth]{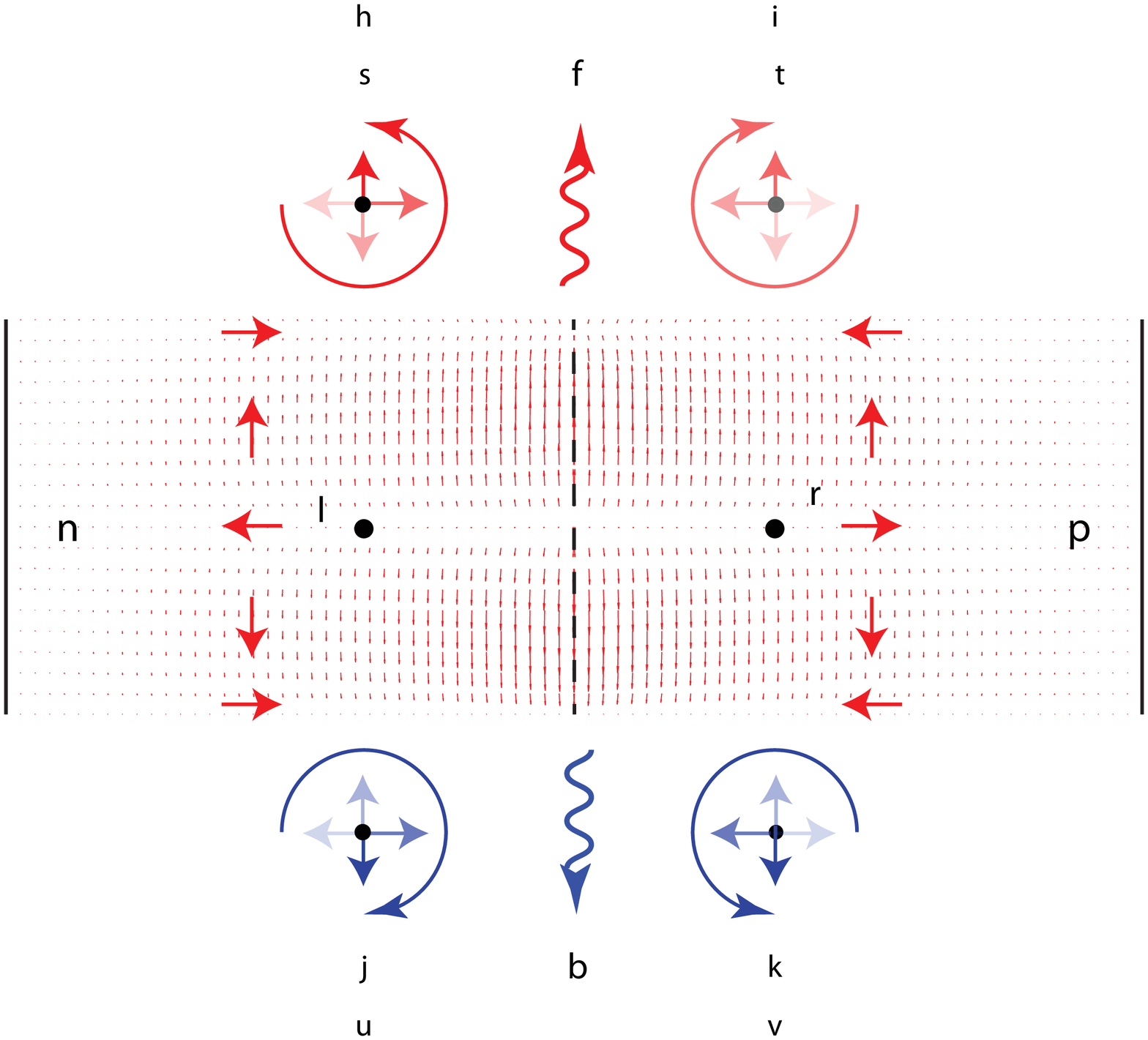}
\caption{Electric field pattern in the plane of graphene for the structure of Fig.~\ref{fig:GrapheneP-N-Junctions_chemical} with $B_0=0$~T. Points R on the right strip sees a clock-wise rotating electric field in the forward direction and a counter clock-wise rotating electric field in the backward direction as the wave propagates. Point L on the left strip sees an oppositely rotating electric field to point R in each direction.}
\label{fig:efield_graphene_plane_B0T}
\end{center}
\end{figure}

\begin{figure}[ht!]
\begin{center}
\psfrag{a}[c][c][0.9][90]{$B_0=0$~T}
\psfrag{b}[c][c][0.9][90]{$B_0=0.2$~T}
\psfrag{c}[c][c][0.9][90]{$B_0=0.5$~T}
\psfrag{d}[c][c][0.9][90]{$B_0=1$~T}
\psfrag{e}[c][c][0.9][90]{$B_0=0$~T}
\psfrag{f}[c][c][0.9][90]{$B_0=0.2$~T}
\psfrag{g}[c][c][0.9][90]{$B_0=0.5$~T}
\psfrag{h}[c][c][0.9][90]{$B_0=1$~T}
\psfrag{m}[c][c][0.9]{mode~$1^-$}
\psfrag{p}[c][c][0.9]{mode~$1^+$}
\includegraphics[width=0.8\columnwidth]{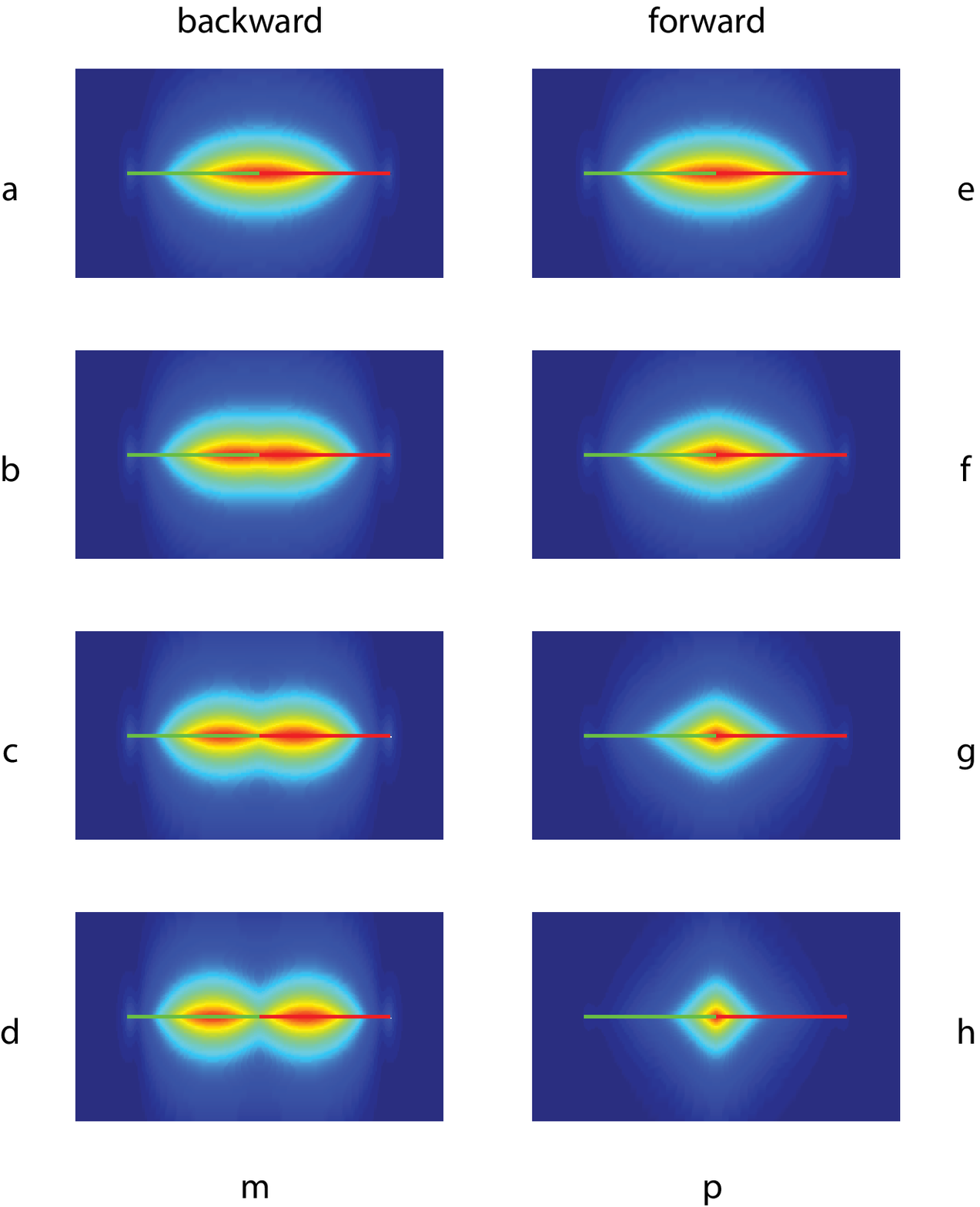}
\caption{Evolution of the mode propagating at the p-n junction (mode 1) for the forward and backward propagations as the magnetic field is increased.}
\label{fig:PN_isolator_mode1_evolution}
\end{center}
\end{figure}

The non-reciprocal property of the magnetically biased structures of Fig.~\ref{fig:GrapheneP-N-Junctions_electrical} and Fig.~\ref{fig:GrapheneP-N-Junctions_chemical}, results from the circular birefringence for the two propagation directions. The electric field pattern of the p-n junction mode in the plane of graphene is shown in Fig.~\ref{fig:efield_graphene_plane_B0T} for the unbiased structure of Fig.~\ref{fig:GrapheneP-N-Junctions_chemical}. As the wave propagates along the strip in the forward direction, point $R$ on the right strip sees a clock-wise rotating electric field, and point L on the left strip sees a counter clock-wise rotating electric field. In magnetically biased graphene, which is characterized by a conductivity tensor $\bar{\bar{\mathbf{\sigma}}} = \sigma_d(\hat{\mathbf{x}}\hat{\mathbf{x}} + \hat{\mathbf{z}}\hat{\mathbf{z}}) + \sigma_o(\hat{\bf{x}}\hat{\mathbf{z}} - \hat{\mathbf{z}}\hat{\mathbf{x}})$, where $\sigma_d$ and $\sigma_o$ are the diagonal and off-diagonal conductivities, respectively, the right and left-hand circularly polarized waves see different scalar conductivities, $\sigma_d + j\sigma_o$ and $\sigma_d - j\sigma_o$, respectively. On the other hand, when a magnetic field is applied, the p-doped and n-doped strips have off-diagonal conductivities with opposite signs. As a result, for the forward propagation both strips see the conductivity $\sigma_d - j\sigma_o$ while for the backward direction the conductivity seen by the wave is $\sigma_d + j\sigma_o$. Therefore, the mode sees different media for different propagation directions, corresponding to different dispersions, as observed in Fig.~\ref{fig:PN_isolator_biased_dispersion}. In the forward direction, the p-n junction mode sees a conductivity with a slightly higher imaginary part and is therefore more concentrated. In the opposite direction, the imaginary part of the conductivity is slightly decreased and the mode becomes less localized. The evolution of mode 1 with increasing magnetic field is plotted in Fig.~\ref{fig:PN_isolator_mode1_evolution} for the forward and backward directions. As the magnetic field is increased, in the forward direction the mode gradually becomes more concentrated on the p-n junction, whereas in the backward direction it moves away from the center.

\section{Conclusions}

The energy concentration and splitting effect allows the realization of non-reciprocal plasmonic devices such as isolators. If the structure is excited at the center, in the forward direction the p-n junction mode is excited, whereas in the backward direction there is only negligible coupling to the backward mode which has little energy at the center. The electrically doped isolator will be very lossy however, as the carrier density is low at the center (the loss is shown in Fig.~\ref{fig:ElectricFieldGatedGrapheneMagnetoPlasmon_b}). In a chemically doped graphene p-n junctions the loss can be mitigated by appropriate doping (the loss is shown in Fig.~\ref{fig:PN_isolator_biased_dispersion_b} for \mbox{$n=p=10^{13}$}).


%
%

%
%
%
%

\bibliography{ReferenceList}

\end{document}